\documentclass[sigconf]{acmart}
\AtBeginDocument{%
  }

\setcopyright{none}
\acmConference[Uploaded to arXiv]{}{}{}
\acmISBN{}
\acmDOI{}

\usepackage{tabularx}
\usepackage{multirow}
\usepackage{xspace}

\usepackage{algorithm}
\usepackage{algpseudocode}
\usepackage{amsmath}

\usepackage{enumitem} % Needed to align itemize to the left (e.g., \begin{enumerate}[leftmargin=10pt] )

\newcommand{\toolurl}{\url{https://models-lab.inf.um.es/modelgraph-ui/}\xspace}
\newcommand{\replicationpackageurl}{\url{https://models-lab.inf.um.es/files/modelgraph-dataset/}\xspace}
\newcommand{\totalmderepos}{7,436\xspace} % Pongo lo que sale en el megamodelo
\newcommand{\totalartefacts}{327,762\xspace}

\newcommand{\modelgraph}[1] {{\textsc{\small\sffamily ModelGraph}}}

\newcommand{\code}[1] {{\small\sffamily #1}}

\usepackage{fontawesome5}
\usepackage{xcolor}
\usepackage{hyperref} % Just load it simply

\hypersetup{hidelinks} 

\definecolor{darkgithubblue}{RGB}{0, 51, 153}

\newcommand{\github}[1]{%
    \href{https://github.com/#1}{%
        \textcolor{black}{\faGithub}% Icon in black
        \hspace{0.3em}% Tiny space between icon and text
        \textcolor{darkgithubblue}{#1}% Project name in dark blue
    }%
}

\usepackage{tcolorbox}

\newcommand{\blacklabel}[1]{%
\tcbox[
    on line,
    colback=black,
    colframe=black,
    coltext=white,
    left=0.10pt,     % Extra padding on the left
    right=0.10pt,    % Extra padding on the right
    top=0.10pt,    % Minimal padding on top
    bottom=0.10pt, % Minimal padding on bottom
    arc=0pt,      % Small rounding
    boxrule=0.10pt,
    fontupper=\small\bfseries\sffamily % Added sffamily for a cleaner UI look
  ]{#1}
}

\begin{document}

%\title{Quo Vadis, MDE?\\
%  Analysing the status of Model-Driven Engineering in public projects\\}

%\title{Building a large scale mega-model of public projects\\The case of the EMF-based artefacts}
% How EMF-based modelling tools are build?

%\title{MDE in the wild\\The case of the EMF-based artefacts}

%\title{How EMF-based modelling environments are build?\\Building a large scale mega-model of MDE public projects}

%\title{\textsc{ModelGraph}: A large scale mega-model of MDE public projects}
%\title{Connecting the Dots: A large scale mega-model of MDE public projects}
\title[Connecting the Models: A Global Mega-model of MDE Projects on GitHub]{Connecting the Models:\\A Global Mega-model of MDE Projects on GitHub}

\author{Jes\'us S\'anchez Cuadrado}
\email{jesusc@um.es}
\orcid{0000-0001-9755-5616}
\authornotemark[1]

\affiliation{%
  \institution{Universidad de Murcia}
  \city{Murcia}
  \state{Murcia}
  \country{Spain}
}

\renewcommand{\shortauthors}{J. S. Cuadrado}

\begin{abstract}
%The Model-Driven Engineering paradigm aims to deliver productivity and quality gains
%by relying on software models throughout the development process. Such models
%are built and managed by developers using domain-specific modelling environments. In theory,
%dedicated technologies like EMF for meta-modelling, ATL and Epsilon for model transformations, Xtext for textual concrete syntaxes, etc.

A key element of Model-Driven Engineering is the construction of domain-specific modelling environments to
improve productivity and quality. In theory,
dedicated technologies like EMF, ATL, Epsilon, Xtext, etc.
would boost the construction of high-quality environments with
a relatively modest effort by chaining the output of one tool to the input of another.
However, there is little empirical evidence of how this idea has fared in reality and 
many open research questions remain, such as how MDE tools are used and combined, 
whether the resulting environments are maintained or not, which tools are used more frequently, etc.

In this paper, we aim to build a foundation for studying how MDE is used in practice.
% by providing a large scale dataset
First, we constructed a dataset by mining \totalmderepos Github projects comprising over 325,000 MDE artefacts.
These artefacts encompass representative Eclipse EMF-related technologies, namely Ecore, Emfatic, OCL, ATL, Epsilon, QVTo, Henshin, Acceleo, Xtext, Emftext, GMF and Sirius.
We also integrated into the dataset repository-level information extracted from the Git repositories and the GitHub API. 
From this dataset, we devised a technique to recover the mega-model of each project in order to represent the relationships between its artefacts.
%This allows the exploration of the configuration of modelling projects and how different types of artefacts are connected.
Then, we built a global mega-model relating the different MDE projects by performing an analysis of
near-duplicates across all artefacts and grouping duplicate artefacts into single nodes and rewiring
the connections. This global mega-model can be used to derive additional information like inter-project dependencies or studying
connected subgraphs of artefacts.
Finally, we propose a number of research questions that could be answered with the provided dataset, which we hope will
foster empirical analysis of how MDE is applied.

%and %we explore the question of how transformation chains are developed.

%Moreover, we perform a near-duplication analysis identifying which files are duplicated among the repositories. From
%this analysis we build a global mega-model relating the different projects by means of the files that they have in common
%and derive additonal information like inter-project dependencies and connected subgraphs.

%To this end, we have analysed 4727 Github projects which contains  more than 175000 modelling artefacts from 4727 Github projects. 

%%   A cornerstone of the Model-Driven Engineering paradigm is the use
%% meta-models, transformations, concrete syntax tools, etc. to construct
%% of domain-specific environments.
%% The use of such environments by developers would boost productivity and quality in software development.
%% Thus, a lot of effort has been devoted to develop and investigate the properties of such tools, yielding well known
%% frameworks and languages like EMF, ATL, Epsilon, Xtext, Sirius, etc. 

  %The Model-Driven Engineering paradigm is based on a few principles like
%the use of meta-models as the pivotal element to build modeling tools,
%the use of meta-tools to...

\end{abstract}


%<concept_id>10011007.10011006.10011060.10011061</concept_id>
%<concept_desc>Software and its engineering~Unified Modeling Language (UML)</concept_desc>
%<concept_significance>300</concept_significance>
%</concept>

%\ccsdesc[300]{Software and its engineering~Unified Modeling Language (UML)}


\received{March 2026}

\maketitle

\section{Introduction}

At the beginning of the century, the Model-Driven Engineering (MDE) paradigm
was considered a promising approach to software development, with the potential to revolutionise the way software was built.
MDE promised to deliver productivity gains by relying on software models throughout the development process.
Such models would be specified and managed by developers using domain-specific modelling environments (either based on general purpose modelling languages like UML
or based on Domain-Specific Languages). Therefore, a key aspect was how to develop these environments by creating (or reusing) model transformations, code generators, validators, simulators, etc.,
using dedicated technologies like EMF for meta-modelling, ATL and Epsilon for model transformations, Xtext for textual syntax, etc.
The end goal was to raise the level of abstraction, and improve automation (e.g., generating code) and quality (e.g., by having domain-specific validation).

%To this end, developers would use existing modelling languages (e.g., UML) or define new domain-specific modelling environments,
%creating (or reusing) model transformations, code generators, validators, simulators, etc., by means
%of dedicated technologies like EMF for meta-modelling, ATL and Epsilon for model transformations, Xtext for textual concrete syntaxes, etc.
% jesusc: No sé si poner esto, creo que no: and/or combining several MDE tools.
%Then, they would create (or reuse) model transformations, code generators, validators, simulators, etc., to enhance DSLs.
% using the most appropriate tooling
%The end goal was to make use of all these artefacts in order to improve automation (e.g., generating code) and quality (e.g., by having domain-specific validation).

Twenty years later, a key question for the software engineering community, and particularly for the modelling community, is whether MDE has fulfilled its promises and, if not, to have insights about the current state and what may have failed. Some works have studied the impact of MDE in industry, showing that MDE can be applied succesfully in narrow domains~\cite{hutchinson2011empirical, liebel2018model}.
%Another way is to study the state of MDE in terms of how its principles have been applied in practice is to analyse open source projects that use modelling technologies.
%This has been studied focusing on some specific types of files, like Ecore meta-models~\cite{de2025analysis, babur2024language}
%and Xtext grammars~\cite{zhang2026development}, and UML models~\cite{romeo2025uml} and BPMN models~\cite{saeedi2025empirical}, but to date are not any works
%that have analyzed complete MDE projects focusing on the relationships betwen the different artefacts that make up a MDE project.
Other works have analysed open source projects to study the usage of specific types of files, like Ecore meta-models~\cite{de2025analysis, babur2024language}
and Xtext grammars~\cite{zhang2026development}, and UML models~\cite{romeo2025uml} and BPMN models~\cite{saeedi2025empirical, compagnucci2021trends}.
However, to date, there have not been any works which have analysed complete MDE projects ``in the wild'' or focused on the relationships between the different artefacts that comprise them.

This work aims at providing the basis to study how MDE tools are built by realising a large-scale analysis of modelling technologies in GitHub.
To perform such analysis we have mined GitHub to identify potential MDE projects and we have collected relevant MDE artefacts. This first phase has yielded a {\em raw dataset}
of~\totalmderepos{}~Github projects and~\totalartefacts{}~artefacts accompanied by stats extracted from the Git repositories and the GitHub projects.
From this dataset, an analysis is performed based on reverse engineering such projects in order to recover their underlying project mega-model (that is, the relationships between the project's artefacts). Then, the recovered mega-models are merged into a global mega-model which represents potential relationships between projects. This process is supported by dedicated algorithms to analyse near-duplicates across artefacts.
%Morever, for each type of artefact specific analysis and stats are gathered.
%In addition, we have performed a semi-automatic classification of the modelling projects in order to provide a better context for the recovered information.
All of this is assembled into a {\em mega-model dataset} which provides several levels of information: project-level mega-model, inter-project relationships and a global mega-model graph. In addition, we have built an interactive tool to explore the data.
% , connected components representing potential relationships across repositories
%In this way, both datasets can be used to perform further analysis and to address key research questions.
We also show the usefulness of the datasets by providing an initial set of research questions which could be addressed.

Altogether, this paper makes two main contributions:
\begin{enumerate}
\item An end-to-end methodology to construct a global mega-model to study the relationships between MDE artefacts. The practical construction of a complete mega-model of publicly available MDE artefacts of such a scale had not been achieved so far.

%\item A method to semi-automatically classify modelling repositories.

\item Two datasets readily available to perform empirical studies on public MDE projects, named \modelgraph. The datasets are supported by a web-based tool to explore the mega-model, as well as databases and a set of APIs to access it. The goal is to enable other researchers easily experiment with this data. This is available at \replicationpackageurl and \toolurl.

%\item We answer a number of RQs based on network analysis. % say which are the main results
\end{enumerate}

% Which are the contributions?

% TODO: Posiblemente poner respuestas a las preguntas

{\noindent\bf Organization}.
The rest of the paper is organized as follows. In Section~\ref{sec:background}, we motivate the paper through a running example, discussing some challenges
and presenting a quick overview of the approach. Then, Section~\ref{sec:raw} describes the construction of the
raw dataset and Section~\ref{sec:megamodel} presents the construction of the global mega-model. Section~\ref{sec:tooling} shows the supporting tools while
Section~\ref{sec:assessment} presents a critical assessment and discusses several research questions which can be answered with this contribution.
Finally, Section~\ref{sec:related} presents related works and Section~\ref{sec:conclusion} concludes.

\section{Motivation and process overview}\label{sec:background}

%% jesusc: Esto ya se ha dicho en la introducción.
%% The goal of the MDE paradigm is to improve the quality and productivity of software development by raising the level of abstraction.
%% In this way, developers build models using existing modelling languages like UML or by developing new DSLs specifically focused on a particular domain (i.e., raise the level of abstraction).
%% These languages are often accompanied by other types of artefacts like model transformations, code generators, textual and graphical editors, validators, interpreters, simulators, etc. (i.e., to improve productivity).
%% In this sense, there are two main flavours of MDE: (1) the use of general purpose modelling languages, like UML and BPMN, which are complemented with custom-made transformations and generators and (2) the definition of new domain-specific modelling languages (DSLs) using a meta-modelling framework plus associated tools to build transformations, code generators, simulators, etc. which are packed in a domain-specific modeling tool. 

%A key approach in MDE is to build domain-specific tooling to allow developers define models and transform and analyse them efficiently.
%A plethora of technologies have been build to realise this goal.

Many technologies have been proposed to enable de construction of domain-specific modelling environments for and with MDE.
A core element is the use of standard meta-modelling frameworks (like EMF, SDF3, MPS Structure Language) which provide a common formalism around which other tools and meta-tools can be built. In particular, in the EMF ecosystem, many meta-tools have been proposed which target different concerns of the construction of modelling environments.
In contrast to integrated language workbenches like MPS or Spoofax, a MDE tool built on top of EMF will use different meta-tools for its construction and possibly custom-made solutions for their integration.
Thus, typical configurations include the use of language creation tools like Sirius or Xtext to define editors, model validation with OCL or EVL, model transformations with languages such as ATL, ETL, QVTo, etc. and code generation
with languages such as Acceleo, EGL templates, etc.
Therefore, a modelling project contains many artefacts conforming to different MDE-specific languages (ej., Ecore, ATL, etc.). From now on we will refer to this kind of files as MDE artefacts.

% Poner más sobre esto
In theory, the combination of these tools would result in high-quality environments, built with moderate effort.
%jesus: esto no lo pongo porque no podemos analizarlo
%which provide the intended improvement in quality and productivity to developers.
However, there is little empirical data to assess whether this claim is true or not and how MDE tools are used to build modelling environments.
This lack of empirical data has left many research questions about MDE open. Some of the questions are
described in Sect.~\ref{sec:usages}, like {\em what has been the evolution of MDE projects in time?}, {\em are MDE projects actively maintained?}, {\em how developers reuse MDE artefacts?}, {\em do developers build transformation chains?}, etc.

This work aims at addressing the gap in the knowledge of how MDE artefacts are used in public projects by realising a large-scale analysis of MDE technologies in GitHub.
%Such analysis have resulted in a large mega-model which is the basis to answer the aforementioned research questions (and possibly others) by the modelling community.
In this work we focus on the EMF because it is widely used and its lack of integration presents a wide range of challenges.
We leave the application of our method to other systems like MPS or Spoofax as future work.

\subsection{Running example}
\begin{figure}[h]
    \centering
    \includegraphics[width=0.50\textwidth]{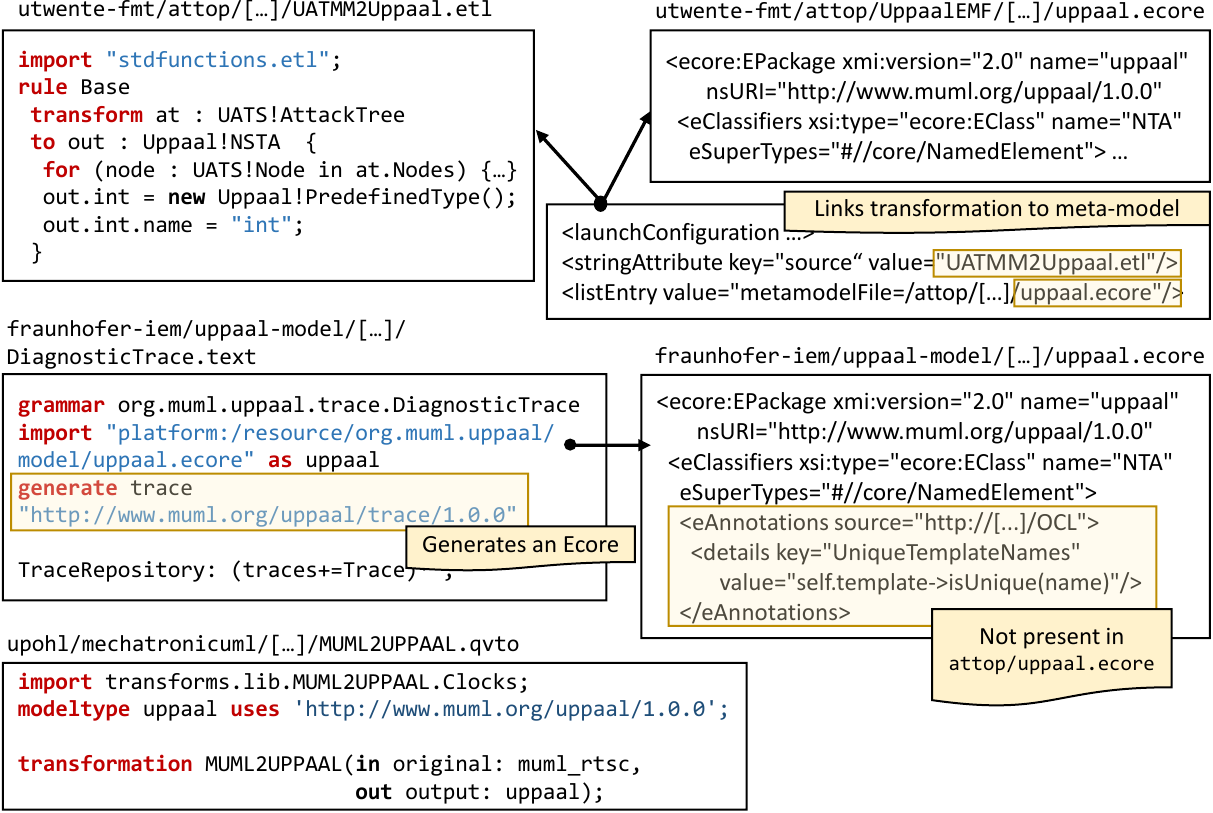}
\vspace{-15pt}
    
    \caption{MDE artefacts in different GitHub projects.}
    \label{fig:running}
\end{figure}

As a running example, let us consider three concrete GitHub projects. \github{utwente-fmt/attop} is a research tool for analysing attack trees, 
\github{upohl/mechatronicuml} which is a fork of a tool for designing and analysing mechatronic systems, and \github{fraunhofer-iem/uppaal-model} which is a wrapper for Uppaal based on EMF.
Fig.~\ref{fig:running} shows some of the files in the projects.
%\code{fraunhofer-iem/mechatronicuml} 
%\code{uppaal-emf/uppaal}
%Project \code{attop} is based on Epsilon for model manipulation (ETL transformations and EGL templates) and targets, among others,
%the Uppaal meta-model.

Project \code{attop} includes the \code{UATMM2Uppal} Epsilon transformation from a custom meta-model (UAT) to the Uppaal model checker. 
To do this, the project made a copy of \code{uppaal.ecore} published in some other GitHub repository. It might be the case that it came from
\code{fraunhofer-iem/uppaal-model} since it to publishes an Uppaal wrapper for EMF (actually, this is a fork of \code{uppaal-emf/uppaal}).
It is worth noting that both files of the Uppaal meta-model are not identical but have evolved independently. In this case,
we show an excerpt of the \code{fraunhofer-iem/uppaal-model} version which includes OCL constraints not present in \code{attop}'s.
At the same time, \code{upohl/mechatronicuml} is based on QVTo, among other technologies, and also targets Uppaal.

Our goal is to recover the relationships between the artefacts that make up MDE projects and to build a global mega-model
to explore them.
However, there are several challenges which make it difficult to systematically perform such an analysis.
%, which we tackle with our approach.
First, the plethora of MDE technologies makes it difficult to build a simple and uniform analysis pipeline since each MDE tool is different and the way it refers to a meta-model or to
other artefacts differs. Moreover, there is generally not a standarized project organization. Thus, we need to build specific analysers for each considered technology and be broad
about the project organizations.
Second, the relationships between artefacts are many times implicit in the sense that are not directly declared in the MDE artefacts.
For instance, an Epsilon program (see example) does not declare its meta-models. Instead, this information needs to recovered from other sources
like configuration files (in the example, an Eclipse launcher contains information about which meta-model corresponds to the transformation).
Third, in MDE projects dependencies are rarely handled systematically, but by copy-pasting and adapting artefacts. In the example, this is evident
by the fact that the \code{attop} project solves it needs to target Uppaal by copy-pasting resources from another project.

\subsection{Process overview}\label{sec:overview}

%In practice, the application of this method results in a large dataset to study different aspects of MDE.
%The process that we have designed to tackle the aformentioned challenges and to produce useful data
%for prospective analysis is summarized in Fig.~\ref{fig:graph_process}.
%There are two main steps: (1)~extract and analyse the raw MDE artefacts and~(2)~build a global mega-model.

To tackle the aforementioned challenges, we propose a systematic method to mine MDE projects, recover their implicit
mega-models and combine them into a global mega-model.
The process we have designed have two main steps: (1)~extracting and analysing the raw MDE artefacts and~(2)~building a global mega-model.

\begin{figure*}[h!]
    \centering
    \includegraphics[width=1.0\textwidth]{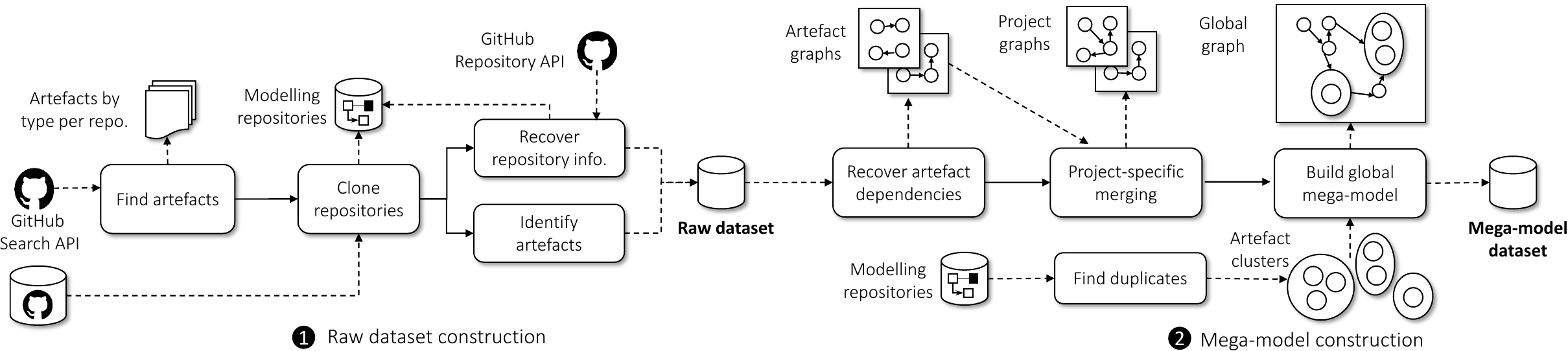}
    \caption{Method to build the mega-model of MDE artefacts from GitHub repositories.}
    \label{fig:graph_process}
\end{figure*}

In the first step, we work at the file and repository levels to extract the raw data. We use the GitHub API
to find relevant MDE artefacts and projects, then we clone them locally to extract the actual list of MDE artefacts
as well as to extract useful statistics. The collected
data is gathered into the {\em raw dataset}.
This is described in detail in Section~\ref{sec:raw}.

In the second step, we apply specific algorithms and strategies to recover the relationships
between the MDE artefacts. The process starts from the raw dataset, considering only those
projects with at least one MDE artefact.
%which provides us with
%a total of \totalmderepos repositories which contain at least one modelling artefact.
It follows three main steps (which are described in detail in Section~\ref{sec:megamodel}):

%The process has two main outcomes: a) the raw dataset containing the set of identified modelling artefacts and repository-level information and b) the mega-model graph.

\begin{enumerate}[leftmargin=*] 
%\item{\bf Artefact identification}. We traverse each repository collecting the modeling artefacts described in Table~\ref{tab:artifactTypes}, organized by .
%  Each artefact is processed independently (the next step) using the actual repository content that was downloaded (see previous section).
%  The rationale is to make sure that the final set of files is gathered from the actual repository contents and not from the GitHub data which might not be completely accurate.
\item{\bf Recover artefact dependencies}. From the raw dataset we traverse the MDE artefacts described in Table~\ref{tab:artifactTypes} and analyse them one by one.
  The goal of the analysis is to recover the direct relationships of each artefact individually. For instance, to which meta-models it conforms to, which artefacts imports, what does it generate, etc. This is discussed in Section~\ref{sec:artefact_analysis}. The output of each artefact analysis is a so called ``artefact graph'' representing its recovered relationships.
  % Therefore, when it finishes the result is a set of unconnected {\em artefact graphs}.
\item{\bf Project-specific merging}. In this step we merge all artefact graphs in a given repository so that we can build a mega-model of the project. The merging strategy is described in Sect.~\ref{sec:merging}.
\item{\bf Build mega-model graph}. Finally, we merge all graphs into a unified, large scale megamodel. This is done by first computing near-duplicates among all artefacts, and then using such duplicates to identify ``merging points'' for the project-specific graphs. This is described in Section~\ref{sec:megamodel_graphs}.
\end{enumerate}

\section{Building the raw dataset}\label{sec:raw}
In this section we describe the methodology followed to build the raw dataset, that is,
a dataset which is solely based on modeling artefacts collected from GitHub, without further
processing to analyse its contents.
% The rationale is...
% This is a threat to validity

%As our dataset is composed of a set of modeling artifacts extracted from existing modeling projects, the resulting dataset becomes a megamodel.
%In the following we will describe first the set of modeling artifacts considered in our study and then the process to build our megamodel.
%As part of the metamodel discovery process, we also describe the relationship extraction process and the corresponding evaluation.

\subsection{Data collection}
The first step was to setup a mining process to crawl GitHub and collect the relevant data. It had the following steps:

\begin{enumerate}[leftmargin=*]
\item {\bf Collecting individual MDE artefacts}.
We started by identifying relevant types of MDE artefacts (see Section~\ref{sec:artifact_types}), for instance Ecore meta-models, ATL transformations, etc. We use the GitHub API to search for files with extensions related to such artefacts of interest. For each artefact type we have defined a GitHub query string which includes a piece of text that always appear in the corresponding artefact. This allows us to filter out most files with the same extension but not belonging to the MDE technical space. For instance, we search for \code{.ecore} files containing the \code{EPackage} text as a hint that it is actually a meta-model. In addition, to overcome the API limit, we split the searches by size ranges, so that we iteratively move a sliding window (e.g., files with sizes between 128Kb and 256Kb) until no files are found in several iterations.

% We have experimented experiments with different searches and the search API do not index the repositories of a user if he or she has not updated any repository within
% Tener en cuenta esto: https://gemini.google.com/share/04d468cfa1d1

% Otra estrategia: si un usuario tiene algún repositorio de modelado, quizás tiene más. Buscar ficheros en esos repositorios => no se indexan

\item {\bf Handling missing repositories}.
  After manual inspection, we found out that some MDE artefacts which can actually be found in GitHub were not in our list. The reason is that GitHub is no longer indexing repositories\footnote{This is briefly described in this GitHub's announcement https://github.blog/changelog/2020-12-17-changes-to-code-search-indexing/.
  %The policy is not clearly stated but after some experimentation we found that only those repositories without any activity (i.e., not being accessed) are not indexed
}
  %although real policy is not explicitly stated.}
with no activity within one year. This means that the results of the previous step may be missing some historical data. To alleviate this,
we perform two actions. First, we have carried out three crawling steps, in February 2022, May 2025 and the last one in March 2026, so that the chances of missing a repository are reduced.
Second, we have also considered the files crawled from GitHub by the MAR dataset~\cite{lopez2022efficient} for Ecore and Xtext which were crawled in 2020, before GitHub introduced the policy and therefore guarantee the inclusion of historical repositories.

 %  From the files, we collect the names of their owning repositories and we download all of them.

  % If you think about it, we don't need to download them, it is enough to query and discover the repositories...

\item {\bf Cloning repositories}.
From the retrieved files, we derive the set of GitHub projects which contain MDE artefacts.
%  To do this we just need to strip the set of owning repositories, \code{user-name/repository-name} from the collected files.
Then, we download all of the repositories to form our dataset. The goal is to make sure that we recover complete MDE projects.
This two-step approach has two advantages:~(1)~it allows us to make sure that we do not miss any artefact not crawled in the first step, and
%For instance, in subsequent analysis we consider build files to discover relationships between artefacts, but it is not practical to crawl GitHub for such files (e.g., there are millions of ANT scripts unrelated to modelling in GitHub).
(2)~by downloading the repository we gather complete information about the project (i.e.,~the Git history).

\item {\bf Discovering artefacts}. All downloaded repositories are traversed collecting the files which fits the artefact type extensions described in Table~\ref{tab:artifactTypes}. In this traversal we perform another, more aggresive content filtering
  step to rule out files which might have been wrongly crawled due to limitations in the GitHub API. For instance, extension \code{cs} used by EMFText is shared with C\# files, which means that any C\# file containing the string ``SyntaxDef'' was included in the original file list. So, we filter out any \code{cs} file not containing several common EMFText tokens like (\code{RULE} and \code{::=}). In addition, we use this pass to collect the creation and last update dates of each artefact. 

\item {\bf Collect repository-level information}.
We use two sources to collect repository-level information:~a)~the GitHub API to retrieve community-related information like project description, stars, forks, etc., and~b)~the Git repository itself in order to collect information like last update, etc.

%% This goes not into another part
%\item {\bf Artefact-specific graph}. For each type of artefact we extract its so-called {\em Recovery graph}. This requires writing specific algorithms and heuristics to detect relationshps between elements. Additional details about this step is given in Section~\ref{sec:artefact_types}.

%\item {\bf Construct dependency graph}. Then, we merge all recovery graphs by using the identifiers of the nodes. The creates a large
\end{enumerate}

\begin{figure}[h]
    \centering
    \includegraphics[width=0.40\textwidth]{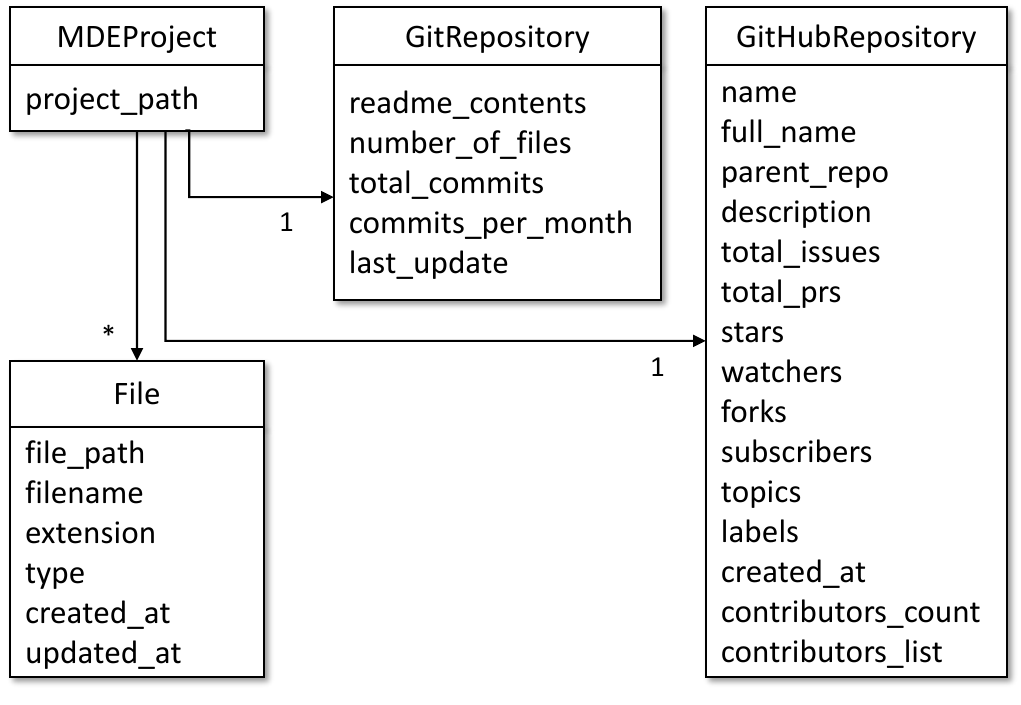}
    \caption{Schema of obtained information.}
    \label{fig:data_schema}
\end{figure}

The final result of this process is a database, named {\em raw dataset}, whose relational schema is shown in Fig.~\ref{fig:data_schema}.
It contains the set of MDE projects along with the list of MDE artefacts (\code{File}). The Git-level information for the repository includes the contents of the README, the number of files in the repository, the date of the last update and some simple stats like the total number of commits. The information related to the GitHub repository contains all the information available from the GitHub API, such as the parent repository (if it is a fork), the number of issues, stars, forks, etc. At the artefact level we record the type of files (which is different from the extension since for example for Epsilon we consider 7 types of extensions) and also the creation date (by looking at the first commit that created the file) and last update. Finally, we keep a private copy of the repositories downloaded in March, 2026 which we use to analyse the contents of the files in the mega-model recovery step.
% hay un fichero raw_process?

\subsection{Artifact types}
\label{sec:artifact_types}

In this work we have considered~12~types of MDE artefacts.
Table~\ref{tab:artifactTypes} summarizes them, including data about number of artefacts, the percentage of ocurrence in projects and the number
of unique files. This data is discussed in Section~\ref{sec:assessment}.

The selection of these artefacts is based on the following criteria: they must be EMF-based artefacts (that is, compatible or directly built over EMF),
they must be well-known tools by the community and it should be possible to find a non-neglible number of artefacts in GitHub and finally, it must be
feasible to download and analyse the artefacts. As an example of discarded artefacts, we have no considered in the current version of the dataset well-known artefacts like VIATRA queries
since the language has undergone non-compatible changes and our manual queries in GitHub have shown that there are little artefacts of each version in GitHub
(i.e., discarded due to the availability criteria). Another example are Xtend templates.
While it is technically possible to include them, it is not clear how to identify automatically which are the source meta-models targeted by the templates (i.e., discarded due to the analysability criteria).

%EMF-based artefacts, well-known tools, availability of artefacts in Github and feasibility of downloading and analysing them.
% The artefacts have been selected on the basis of the popularity of the modelling tools that implement them since our aim is to be able to collect a number of artefacts signficant enough of each type.

We have also considered two types of ``build artefacts'': Eclipse launchers and Ant builds. They are useful in the mega-model recovery phase
to identify artefact relationships which are encoded as configurations instead of directly in the artefacts.
% Esto en la discusión.
%The artefact count in the table shows the total number of files, but most of them do not
%contain relevant information and it is the responsibility of the mega-model recovery phase to analyse this fact.

%  select count(*), type from files where type in ('acceleo', 'atl', 'ecore', 'emfatic', 'emftext', 'epsilon', 'henshin', 'ocl', 'qvto', 'sirius', 'xtext', 'gmf') group by type;

% TODO: List the set of files
\begin{table}[]
  \caption{Artifacts types considered plus some statistics. Column \textsc{Count} shows the total number of
    artefacts of each type, \textsc{Project ocurrences} reflects the percentage of projects with at least one artefact of the given type
  and \textsc{Unique} indicates the total number of unique files (i.e.,~after deduplication).}
  \label{tab:artifactTypes}
  \small
  \begin{tabularx}{\columnwidth}{lXrrr}
    \textsc{Tool}  & \textsc{Purpose} & \textsc{Count} & \textsc{Project}    & \textsc{Unique} \\
                   &                  &                & \textsc{Ocurrences} &  \\
    \toprule

\multirow{1}{*}{Acceleo} & Codegen & 17,664 & 8,98\% & 8,298 (46.98\%) \\ \midrule
\multirow{1}{*}{ATL} & Transformation & 9,850 & 7,25\% & 2,460 (26.60\%) \\ \midrule
\multirow{1}{*}{Ecore} & Metamodel & 199,025 & 93,37\% & 31,099 (15.27\%) \\ \midrule
\multirow{1}{*}{Emfatic} & Metamodel & 2,402 & 5,98\% & 1,475 (62.90\%) \\ \midrule
\multirow{1}{*}{Emftext} & Syntax & 1,442 & 1,00\% & 415 (93.05\%) \\ \midrule
\multirow{1}{*}{Epsilon} & Transformation & 14,234 & 7,42\% & 6,608 (49.04\%) \\ \midrule
\multirow{1}{*}{GMF} & Syntax & 1,280 & 5,62\% & 656 (51.25\%) \\ \midrule
\multirow{1}{*}{Henshin} & Transformation & 52,158 & 1,51\% & 7,755 (14.87\%) \\ \midrule
\multirow{1}{*}{OCL} & Validation & 8,801 & 4,94\% & 4,360 (50.29\%) \\ \midrule
\multirow{1}{*}{QVTo} & Tranformation & 8,380 & 3,63\% & 2,109 (25.29\%) \\ \midrule
\multirow{1}{*}{Sirius} & Syntax & 4,307 & 8,85\% & 1,415 (32.87\%) \\ \midrule
\multirow{1}{*}{Xtext} & Syntax & 8,219 & 41,41\% & 3,995 (48.93\%) \\ \midrule

\multicolumn{2}{r}{\bf Total artefacts } & 327,762 & & \\ \midrule\midrule

    \multirow{1}{*}{Eclipse}&	 Launcher & 29,077 &- & - \\
    \multirow{1}{*}{Ant file}	     &   Build & 16,781 & - & - \\ \midrule
    \multirow{1}{*}{}  &  {\bf Total configs} & 45,858 & & \\ %\midrule %\midrule

    % \multirow{1}{*}{Maven}	        & pom.xml   & Build & XXX \\
 % MWE2
    \bottomrule
  \end{tabularx}
\end{table}

%.javajet|jet|6828
%.jet|jet|1810
%.xml|maven|128648
%.mps|mps|3598
%.mwe2|mwe2|7910
%.ocl|ocl|8801
%.qvto|qvto|8380
%.odesign|sirius|4307
%.sdf3|spoofax|58
%.xtend|xtend|25941
%.xtext|xtext|8219

%.launch|eclipse-launcher|29077

% Other types of files that we could consider:
% - Kermeta
% - OclInEcore?
% - UML files or known DSL files?

\section{A mega-model of MDE in GitHub}\label{sec:megamodel}

The raw dataset is already a valuable resource to study different aspects of MDE (see Sect.~\ref{sec:usages}). However,
it only provides an isolated view of each artefact or repository-level statistics. Hence, we are interested in identifying relationships between
artefacts in order to have a more complete view of the organization of the MDE projects.

To this end, we have built a {\em mega-model} of the projects available in the raw dataset.
A mega-model is defined as a high-level, "model of models", so that it characterizes the relationships between various models, metamodels, and artifacts within a system~\cite{bezivin2004need}.
This section describes the process that we have devised to recover a large scale mega-model
of MDE artefacts found in GitHub projects. % identified as described in the previous section.

%% \begin{figure*}[h]
%%     \centering
%%     \includegraphics[width=0.75\textwidth]{figures/graph-process.pdf}
%%     \caption{Methodology to build the mega-model of MDE artefacts from project repositories.}
%%     \label{fig:graph_process}
%% \end{figure*}

\begin{figure*}[h]
    \centering
    \includegraphics[width=0.95\textwidth]{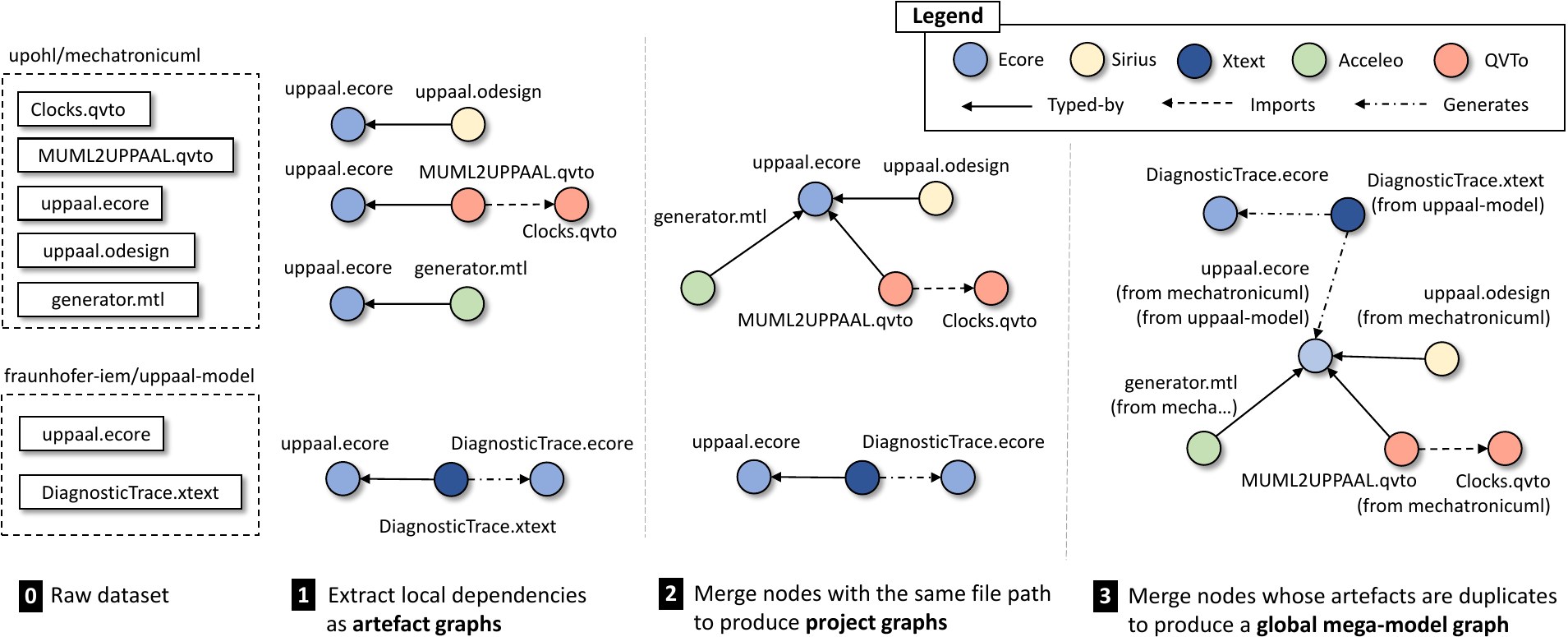}
    \caption{Mega-model recovery applied to artefacts from \github{upohl/mechatronicuml} and \github{fraunhofer-iem/uppaal-model}.}

    \label{fig:process}
\end{figure*}

As a concrete example to guide the explanation, Fig.~\ref{fig:process} shows the process in detail for some of the files of the files
shown in Fig.~\ref{fig:running} (with small modifications to illustrate all cases).

%% For instance, centering the attention into the Uppaal meta-model it is possible to identify projects which perform
%% formal analysis in different domains. The inter-project view allows us to identify such projects. For instance \code{utwente-fmt/attop}
%% verifies attack trees, \code{sillymoi/remes-ide} models embededded software, \code{fraunhofer-iem/mechatronicuml} designs and analyses mechatronic systems, etc.

%{\bf TODO: Explain the example}

%\subsection{Individual artefact analysis}
\subsection{Recovery of artefact dependencies}
\label{sec:artefact_analysis}

In this step we analyze each MDE artefact with the goal of recovering its dependencies. For each artefact we want to obtain
a small graph, named {\em artefact graph}, in which the artefact is the central node and it has edges pointing to the artefacts it depends on.
This is illustrated in Fig.~\ref{fig:process}~(label~\blacklabel{1}). For instance, \code{MUML2UPPAAL.qvto} is the central node of its artefact graph and it has dependencies
with files \code{uppaal.ecore} and \code{Clocks.qvto}. A similar strategy is followed for \code{uppaal.odesign}, \code{generate.mtl} and \code{DiagnosticTrace.xtext}.

Technically, to analyze each artefact we define the so called {\em Artefact inspector}. An inspector takes an artefact file and
generates an {\em artefact graph}. Table~\ref{tab:injectors} describes the inspectors that we have implemented and the strategies
that are used to recover its dependencies.
%(focusing on the meta-model, since explicit imports are more straightforward).
These strategies are explained in the following.

%% We consider three main types of dependencies:
%% \begin{enumerate}
%% \item Typing relationship w.r.t. to the meta-model. In the example, \code{uppaal.odesign} is typed by \code{uppaal.ecore}.
%% \item Import dependencies to reuse functionalities defined in other files. In the example, \code{MUML2UPPAAL.qvto} imports file \code{Clocks.qvto}.
%% \item Artefact generation, when one artefact generates another artefact (e.g., a conversion from Emfatic to Ecore) which may appear in the
%% repository explicitly or not.
%% \end{enumerate}
% For the sake of brevity, we focus the explanation in the meta-model typing relationship since import relationships are handled similarly.

\subsubsection{Dependencies recovery}
There are two main types of dependency relationships upon which the other relationships are built: typing relationships and import relationships.

\medskip
{\bf\noindent Recovering typing relationships}.
All MDE artefacts refer to one or more metamodels directly or indirectly.
In the simplest case the artefact includes an explicit reference to the meta-model (e.g., Xtext has the \code{import} statement
to refer to a meta-model). Sometimes the meta-model is generated from the artefact (e.g., Xtext provide the \code{generate} statement
to indicate this fact). Both cases are shown in Fig.~\ref{fig:running} (file \code{DiagnosticTrace.xtext}).
However, some modelling tools use a ``dynamic linking'' strategy in the sense that meta-models are only known are runtime. For instance, in Epsilon the transformation files
do not contain a reference to their meta-model(s). The advantage is that the transformation is not  bound to a specific meta-modelling framework and modelling drivers
can be used to run the same transformation with different model formats~\cite{zolotas2020bridging}. The downside is that the actual meta-model needs to be passed as part of the run configuration.
Finally, there are tools that that support both explicit meta-model references and dynamic linkage of the metamodel (e.g., ATL and Sirius).
% In the case of models serialized as XMI, the link to their meta-model is typically by means of the meta-model URI.
Given these scenarios, we identify four possible meta-model recovery strategies:
% Poner un dibujo de cada caso

\begin{itemize}[leftmargin=*]
\item {\bf Extract direct references}. This is the most straightforward approach, which is available for ``statically typed'' artefacts.
  For each type of tool we have to figure out the way it encodes a reference to its meta-model.
  For instance, QVTo uses the notion of model type and we need to be able to resolve references to Ecore files relative to the current transformation,
  but also consider the case of an Ecore file pointed by its URI.
  %For instance, in Henshin the meta-model can be found as an \code{href} tag at the XML level.
\item {\bf Imitate generation behaviour}. For tools which automatically generate the meta-model, we need to emulate this behaviour in the recovery process.
In particular, for Xtext we know that the meta-model is automatically placed in the \code{model/generated} folder.
\item {\bf Inspect configuration files} like Ant's \code{build.xml} or Eclipse launchers to know which meta-model is used at runtime.
  Many times the repositories contain the configuration files that developers have used to launch their transformations at development time.
  We analyse patterns typically used by ATL and Epsilon developers in order to identify the meta-models of a transformation.
  %This is useful for tools like Epsilon in which meta-models are only provided at runtime.

%jesusc: Esto no lo hacemos, habria que llevarlo a la parte de threats y trabajo futuro
%\item {\bf Inspect programmatic wrappers}. Once a modeling language environment is finished, developers sometimes wrap the configuration and invocation of the different
%  elements of the environment (i.e., editors, transformations, etc.) into an Eclipse plug-in. Such implementations could also be inspected to extract
%  information about typing relationsihps.

\item {\bf Compare the footprint of the transformation.} For ``dynamically typed'' artefacts it is possible to extract an approximate meta-model footprint of a transformation~\cite{lara2019automated},
by analysing its references to meta-classes. Then, we match such footprints with the Ecore meta-models found
in a project and the EMF built-in meta-models. If the 95\% of the classes in the footprint can be found in a meta-model, we heuristically select it as the
meta-model of the artefact.
\end{itemize}

% TODO: Habría que poner un ejemplo de cada cosa

\medskip
{\bf\noindent Recovering import relationships}.
The recovery of import relationships is simpler since it is typically possible to find a static reference to the imported artefact in the artefact definition. Nevertheless, there
are cases in which such references do not point exactly to project artefacts. This is described in more detail in Sect.~\ref{sec:nodes}.

\subsubsection{Artefact relationship types}
\label{sec:relationships}
% Esto es una subsección
%Esto ponerlo en
%Extracting relationships of the MDE artefacts within a project is a particularly complex and error-prone step because there is a plethora of technologies and each one has its particularities, as has been described above.
%{\bf HABLAR DE QUE HAY RELACIONES QUE SON HEURÍSTICAS}

%{\bf DOCUMENTAR LOS TIPOS DE ARCOS} % luego los arcos se mezclan, supongo
As explained, in an artefact graph, a directed edge represent a typing or import dependency between an artefact (the dependent) and another artefact (the dependency).
In addition, we attach metadata to each edge according to its role with respect to the referenced artefact. So far we consider the following types of relationships within artefacts in a project:
\begin{itemize}[leftmargin=*]
\item {\em Typed-by}. An artefact is typed by some meta-model.
\item {\em Imports}. An artefact imports another artefact as library or helper.
\item {\em Generates}. An artefact generates another artefact (e.g., a conversion from Emfatic to Ecore) at development time and such a target artefact may appear in the repository explicitly or not.
\item {\em As-input}. An artefact uses instances of the other artefact as input. This is used to identify the input meta-models of a given transformation artefact.
\item {\em As-output}. Similar to {\em as-input}. An artefact generates instances of the other artefact as output.
\end{itemize}

Please note that these relationships are not exclusive. For instance, an Xtext file maybe ``typed by'' an Ecore meta-model and such relationship is also labelled as ``generated'' if the meta-model
is generated by Xtext instead of using an existing one.

%% \subsubsection{Recovery graph generation}
%% \label{sec:graph_generation}

%% The strategy to generate the recovery graph is as follows:
%% \begin{enumerate}
%% \item For each relevant file in the dataset, we identify the corresponding artefact inspector according to its file extension.
%% \item When the file represents an artefact, the inspector generates a node representing the arterfact. When the file is a configuration file (e.g., an Eclipse launcher)
%% the inspector try to identify which artefact are involved (typically transformations).
%% \item The inspector uses the strategies described above to identify the corresponding meta-model and potential import relationships.
%% \item The identified meta-models and imported elements generate new nodes in the graph which are linked to the central node by edges annotated with the meta-data explained in Sect.~\ref{sec:relationships}.
%% \end{enumerate}

%% However, the process is more complex because a common situation is that an artefact refers to files that cannot be found reliably in the repository.
%% This may happen because of several reasons:

\subsubsection{Artefact nodes}
%\subsubsection{Dealing with broken dependencies}
\label{sec:nodes}
Each artefact file is mapped to a node in the graph as it is processed when its artefact graph is created.
%by the corresponding inspector.
However, as noted before, such artefact may refer
to other files (dependencies) which may or may not by present in the repository.
Our approach to deal with cases in which a dependency cannot be recovered is to generate a dependency node in the graph even if it does not exist physically in the repository.
The rationale is to record such information in the mega-model as well for further analysis (e.g., analyse broken projects). There are four cases:
% In this way, each generated node representing a dependency is annotated with one of the following categories:

%\begin{itemize}
%\item The dependency file is in the repository and a new node and the corresponding edge can be created.
%\item The file is not actually present (and thus the project is broken).
%\item The file is present but the artefact does not correctly point to it.
%\item The file is present but our recovery algorithm cannot faithfully recover it (e.g., it uses runtime information, environment variables, etc. to construct the path)
%\end{itemize}

\begin{itemize}[leftmargin=*]
\item {\em Exists}. The dependency file is found in the repository in its expected path.

\item {\em Generated}. A referred file does not actually exist in the filesystem but it is straightforward to generate it.
  An example is the Ecore file \code{DiagnosticTrace.ecore} generated by \code{DiagnosticTrace.xtext}, which may not exist in the repository.
%  An artefact graph to represent this relationship is generated, and later merged with the result of \code{uppaalXML.xtext}.
\item {\em Heuristic}. This means that an artefact refers to a file in a way that it is not possible to identify the exact
  file in the repository or that the reference is ambiguous.
  This happens when the file is present but our recovery algorithm cannot faithfully recover it. For instance, in an Ant file for configuring an Epsilon transformation
  a file path may be constructed using runtime information like environment variables (e.g., something like \code{\$\{projectDir\}/MyMetamodel.ecore}). Since the environment variable
  is only available dynamically, we may not know the actual file. However, if we find a file \code{MyMetamodel.ecore} in the repository we can heuristically assume that the
  transformation is actually referring to this file.
% TODO: Try to find another example
\item {\em Missing}. A file is expected but it cannot be found and the heuristic rules cannot be applied. In practice this means
that the project is broken (e.g., it may not be possible to make it work because it lacks some artefacts).
% \item Unexpected -> No sé cómo explicar esto. Es un fallback cuando hay algo que no se controla.
\end{itemize}

% What happen if we have more than one witness for the same artefact

At the end of this process, each artefact file in the raw dataset has an associated artefact graph encoding its relationships with
other artefacts in the repository.

\begin{table}[]
  \small
\caption{Artefact injectors currently implemented and meta-model recovery strategies used.}
\label{tab:injectors}
\begin{tabularx}{1.0\columnwidth}{lll}        %& Antes ponía Analysis
\textsc{Technology}      & \textsc{Analysis} & \textsc{Meta-model recovery} \\ \toprule
\multirow{1}{*}{Acceleo} & Custom parser     & Direct                       \\
\multirow{1}{*}{ATL}     & AST analysis     & Direct, Footprint, Configurations \\
\multirow{1}{*}{Ecore}   & Load model        & N/A                          \\
\multirow{1}{*}{Emfatic} & Custom parser     & Generated                    \\
\multirow{1}{*}{Emftext} & Custom parser     & Direct                       \\
\multirow{1}{*}{Epsilon} & AST/Custom parser & Footprint, Configurations    \\
\multirow{1}{*}{GMF}     & XPath             & Direct                       \\
\multirow{1}{*}{Henshin} & XPath             & Direct                       \\
\multirow{1}{*}{OCL}     & Custom parser     & Direct                       \\
\multirow{1}{*}{QVTo}    & AST analysis      & Direct                       \\
\multirow{1}{*}{Sirius}  & XPath             & Direct, Footprint            \\
\multirow{1}{*}{Xtext}   & AST analysis      & Direct, Generated            \\
\toprule
% Xtend
\end{tabularx}
\end{table}

\subsection{Project-specific graphs}
\label{sec:merging}
% Explicar "el rationale" para tener esto en varios pasos
%An {\em artefact graph} provides a view of the relationships of just one modelling artefact with other artefacts it depends on (e.g., meta-models, generated files, imported modules, etc.).
The artefact graphs are our intermediate representation to encode dependencies.
From this, we are interested in building a larger graph which explicitly reflects the transitive relationships between the recovered artefacts. For instance, in Fig.~\ref{fig:process}~(label~\blacklabel{1})
we can observe that the recovered graphs for project \code{mechatronic-uml} (for the artefacts \code{uppaal.odesign}, \code{MUML2UPPAAL.qvto} and \code{generator.mtl}) actually depend on the same file, the \code{uppaal.ecore} meta-model.
Hence, it makes sense to combine the different artefact graphs found in a project based on their overlapping artefacts.

Our strategy is therefore to merge nodes that refer to the same artefact.
In the example, this means merging all nodes that refer to the \code{uppaal.ecore} meta-model and rewiring the corresponding edges, obtaining the graph shown in Fig.~\ref{fig:process}~(label~\blacklabel{2}).
A real example of a project graph is shown in Fig.~\ref{fig:project_graph}.

%Our strategy is to merge nodes that refer to the same artefact. This is exemplified in Fig.\ref{fig:process}. In project \code{fraunhofer-iem/mechatronicuml} there are several artefacts which
%are typed against the \code{uppaal.ecore} meta-model. To obtain a consolidated graph we merge those nodes that refer to the same artefact. In the example, this means merging all nodes that refer
%to the \code{uppaal.ecore} meta-model, obtaining the graph shown in Fig.\ref{fig:process}(XXX).

Technically, what we do is to derive an identifier for each artefact using the file path of the artefact relative to the project. This allows us to make sure that merged nodes refer to the same artefact. The case of merging meta-model nodes is a bit more complex because a meta-model can be referenced by its URI or by its file path. When it is referenced by its file path, we apply the regular merging process. If it is referenced by a well known URI\footnote{We use the list of meta-models URIs built-in in the Eclipse modelling package.} (e.g., Ecore, UML, etc.) we just use this URI as the node identifier. If the URI is not known, we use an internal index, which maps URIs to meta-model files, to know which file in the project corresponds to such an URI and make the artefact depend on the actual file.
Another caveat is merging {\em heuristic} nodes. When an edge is created whose target is an node created from an heuristic node two actions are performed: the node is labelled with {\em exists} since the file
actually exists in the repository and the edge is annotated with {\em heuristic} to signify that the recovered relationship may be a false positive.

The merging process results in a set of graphs which describe the relationships between the artefacts in each GitHub repository.
We call them ``project-specific graphs''. These graphs can be used, among other purposes, to explore the configuration of MDE projects.

%After this point we have ``project-specific graphs'' which shows information about how a project is organized.

%% Therefore, to address potential issues related to the uncertainty of whethere we have been able to implement adequate recovery algorithms and heuristics, we have followed this methodology:
%% \begin{verbatim}
%% run mega-model recovery

%% for each artefact type:
%%   identify suspicious node (e.g., isolated node)
%%   if exists suspicious node
%%      debug and review extraction code
%%      if node can't be fixed (e.g., it is legit)
%%         add node to list of non-suspicious nodes

%% sample 25% of the relationships
%% if they are ok, then ok, else fix and re-run

%% \end{verbatim}

%% \begin{table}[]
%% \caption{Types of graphs}
%% \label{tab:graphs}
%% \begin{tabularx}{\columnwidth}{llp{4.5cm}}
%% \textsc{Graph} & \textsc{Scope}                          & \textsc{Description} \\
%% \multirow{1}{*}{Artefact}   & Single file                & Describe dependencies of a single file \\
%% \multirow{1}{*}{Project}    & Repository                 & Relationships between artefacts in a single project \\
%% \multirow{1}{*}{Mega-model} & Global                     & Relationships of all artefacts in GitHub using duplicates a join point \\
%% \multirow{1}{*}{Inter-project} & Global                  & Relationships between GitHub projects \\
%% \toprule
%% % Xtend
%% \end{tabularx}
%% \end{table}

% What happens if a project refers to something outside its root with "../". TODO: Check

\subsection{Mega-model graphs}
\label{sec:megamodel_graphs}
Once we have constructed a graph with the relationships between artefacts in the repositories, in this step we compute a global graph which connects the different repositories.
To tackle this, a key observation is that in current MDE technology there is not a package manager (e.g., Maven, NPM, etc.) to facilitate artefact sharing. Instead, developers typically copy-paste artefacts as way
of reusing and adapt them. This is shown in Fig.~\ref{fig:process}, where project \code{uppaal-model} publishes a meta-model of the Uppaal tool along with a textual
syntax (\code{DiagnosticTrace}) to describe the trace of the model checker. Then, project \code{mechatronicuml} reuses the meta-model by making a copy and possibly some
small adaptations. This means that we cannot use techniques devised to map software artefacts globally~\cite{DBLP:journals/tse/DecanM21} since the projects relationships are not explicit.

To recover currently implicit project relationships, we adopt the following assumption: {\em if a repository contains a duplicate or near-duplicate artefact from another repository,
that artefact would have been shared if a ``modeling package manager'' were available.}
To this end, we compute near-duplicates of all of the MDE artefacts by creating specific adaptations of the algorithm proposed by Allamanis~\cite{allamanis2019adverse} to the different artefact types.
The algorithm outputs a set of clusters so that each cluster contains the artefacts that are similar.
In practice, for textual languages we use the language tokenizer. For XMI-based artefacts we extract the identifiers. We could not apply the approach for Acceleo, Emfatic, Emftext and OCL because we could not
access their tokenizers programmatically. For these cases we relied on computing the MD5 sum of the files to identify exact duplicates.

In this way, the construction of the global mega-model takes as input the set of project graphs and the duplication clusters and
outputs a single graph in which artefact nodes appearing in a duplication cluster (i.e., this means that the artefact has duplicates in other repositories) are replaced by a single node representing the cluster and the edges are redirected. These nodes are called {\em artefact groups}.
%proceeds as follows. For artefact in the project graph, we look it up in the duplication clusters. If it can be found, we annotate it

\begin{figure}[h]
    \centering
    \includegraphics[width=0.40\textwidth]{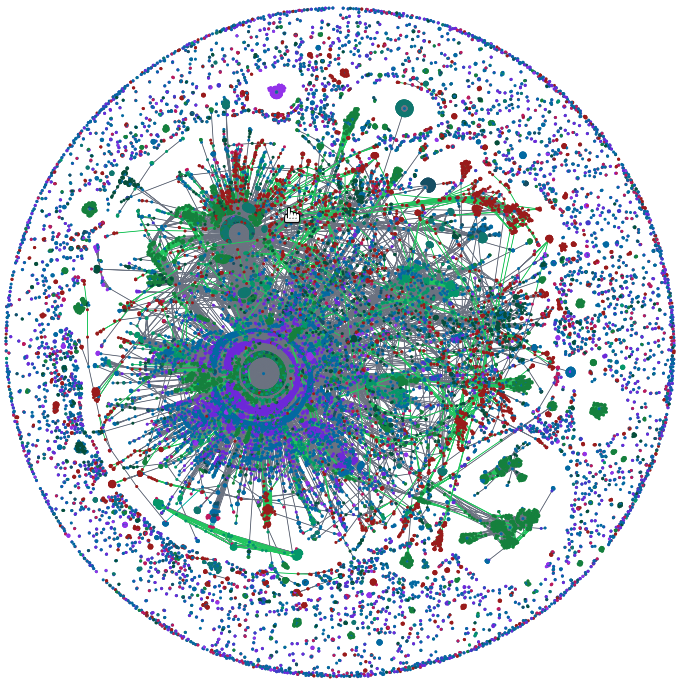}
    \caption{Megamodel visualized with our exploration tool.}
    \label{fig:megamodel}
\end{figure}

In this way, our mega-model makes reuse relationships between projects explicit by representing common artefacts with a single node.
In Fig.~\ref{fig:process}(label~\blacklabel{3}) the element that is duplicated in both projects is \code{uppaal.ecore}. In the mega-model, both nodes are merged into a single node
(the artefact group) and the edges are rewired to point to the merged node. Fig.~\ref{fig:megamodel} shows the mega-model of our dataset, in which it is possible to observe
several large hubs (Ecore, UML) and many other connected components.

% Ver qué pasa con las URIs de los meta-modelos
% We merge all artefact graphs

%From this {\em mega-model} graph, it is possible to derive other graphs which show other perspectives of the global relationships. In particular, we have currently derived two additional graphs: {\em inter-project graph} and {\em connected components graph}:

From this {\em mega-model} graph it is possible to derive other useful graphs or pieces of information. In particular, in this work we also compute the so called
{\em inter-project graph}. This graph is intended to represent dependencies between projects, so that a project A depends on another project B if it has an artefact
that is defined in project B. We take advantage of the artefact groups stored in the mega-model to derive this graph.
%The mega-model graph already represents artefacts potentially shared between projects by storing them in a common node (i.e., duplication node is group of artefacts deemed as duplicates) so that
%other nodes that depend on any of the files stored in the duplication node points to it. There is not a unique way to build this graph.
A simple approach is to consider
that a project is related to another if it contains a duplicated element from such a project. This approach assumes that in a artefact group we do not know which one is the original, and therefore
generates an edge among all projects which contain duplicates.
To make the graph construction more precise, we take advantage of the temporal information recorded in the raw dataset (attribute \code{createdAt} in Fig.~\ref{fig:data_schema}).
We apply the heuristic that, in an artefact group, the original artefact is the older one. In this way, a project depends only on the project that contains the original artefact.

This graph is useful to identify related projects. For instance, if we are interested in understanding which projects
perform formal analysis using Uppaal, we can find a promising seed project by looking up the \code{uppaal.ecore} or similar artefacts. Then, from a project defining such artefact (i.e., \code{uppaal-model}) we can discover projects like \code{utwente-fmt/attop} which verifies attack trees, \code{sillymoi/remes-ide} to model embededded software, \code{fraunhofer-iem/mechatronicuml} for designing and analysing mechatronic systems, etc.

\section{Dataset availability and tooling}\label{sec:tooling}

%In the previous sections we have described the process to derive a global mega-model of modelling artefacts.
The tangible product of the process described below consist of two datasets: the raw dataset and the mega-model graph. Both datasets
are important contributions to improve our knowledge about how MDE has fared in public repositories.
The complete dataset, which we name \modelgraph, and the source code to reproduce the complete process is available at \replicationpackageurl. We also provide a web tool to facilitate the exploration of the dataset at \toolurl.

%In this section we present the tooling that we have built for the modelling community to be able to profit from
%these dataset.

\subsection{Datasets and APIs}
The datasets consist of two SQLite databases plus a Java API to reconstruct the different graphs and access the artefact and repository information.
The raw database has the schema shown in Fig.~\ref{fig:data_schema}.
We store the mega-model in a relational database for convenience. In practice, we load it as a JGraphT\footnote{https://jgrapht.org/} graph which allows
users to navigate the graph (e.g., obtain outgoing edges of a node) and apply algorithms very easily (e.g., find connected components).
%In addition, we provide a REST API to query the raw dataset and the graph.

\subsection{Exploring tool}
\begin{figure}[h]
    \centering
    \includegraphics[width=0.5\textwidth]{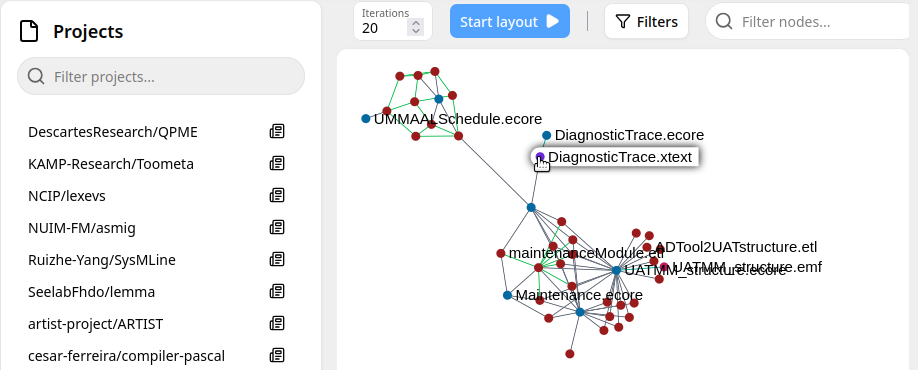}
    \caption{Screenshot of the tool, showing the project graph of \github{utwente-fmt/attop}.}
    \label{fig:project_graph}
\end{figure}

To facilitate the exploration of the provided data we have built an interactive tool. The tool was initially intended to allow us to inspect and debug
our results, but we finally decided to polish it and make it available as part of the contribution. It can be used both locally or remotely at \toolurl.
Fig.~\ref{fig:project_graph} shows the project of \github{utwente-fmt/attop}.
The tool provides the following features: It displays statistics about the artefacts, explore projects graphs (with facilities search for projects),
explore the inter-project graph and the exploration of the global mega-model. It also provides facilities for filtering nodes and edges by type.

%Fig.~\ref{fig:megamodel} shows the mega-model as shown by the tool.

%% The tool has the following features:
%% \begin{itemize}
%% \item Perform queries directly over the repository information.
%%   % For instance:
%%   %  To extract all epsilon projects which may have ant files associated
%%   %    select * from files where type = 'epsilon' and project_path in (select distinct project_path from files where type = 'ant');

%% \item Perform queries directly over the mega-model, and obtain
%%   nicely formatted results.
%% \end{itemize}

%\subsection{Data Availability}

% https://anonymous.4open.science/r/modelgraph/README.md

\section{Discussion}\label{sec:assessment}

In this section we assess the dataset from different perspectives. First, we discuss its limitations and potential shortcomings.
Then, we present a discussion regarding the contents of the dataset. Finally, we discuss potential usages.

% O podría ser simplemente discussion and threats to validity
\subsection{Limitations and threats}
To use the dataset effectively it is important to take into account its limitations and how they may affect (or not) its use.
We have classified these aspects in six categories which are described next.
% Ver si decir internal validity?

\smallskip
{\bf\noindent Completeness}.
Regarding the completeness of the dataset, as we have explained, we may have missed some repositories which do contain MDE artefacts, in particular those created between 2020 and 2025 which have not had any activity
and are not currently indexed by the GitHub search API. We have tried to alleviate this by also cloning repositories from the MAR dataset. In total, the MAR dataset added about 1,000 repositories which are not
available through the GitHub API.  

\smallskip
{\bf\noindent Focus on Eclipse technologies}. Another issue related to completeness is that the dataset is currently focused on modelling technologies around the EMF ecosystem.
We have tried to cover a wide range of technologies in the ecosystem, in particular those with more usage. In future versions we could include other EMF-based technologies not covered so far
like Kermeta, Viatra, etc. On the other hand, there are other technical spaces worth considering like MPS, Spoofax or Monticore. In future work we plan to replicate our approach for these spaces
and compare the resulting mega-models.

\smallskip
{\bf\noindent Correctness}. We have manually implemented inspectors for many types of MDE artefacts. A potential threat is that we might be missing some features of the languages or having made some implementation
mistake. Given the sheer amount of processed artefacts, our approach to ensure the correctness of the implementation has been to sample a number of projects after each major implementation change and manually check the correctness of the graph, fixing bugs as needed. 
%As a correctness metric we use...

%There are sources of uncertainty. For instance, we can have an Acceleo transformation which uses a URI, but this URI is provided by two or more files in the project. This is an example that was ``well-resolved'' because the URI changed with the meta-model version, but this is might not be the general case:
%\begin{verbatim}
%./mechatronicuml/PlatformModeling/org.muml.pm.hardware.migrator/model/hardware-1.0.ecore
%./mechatronicuml/PlatformModeling/org.muml.pm.hardware.migrator/model/hardware-1.1.ecore 
%\end{verbatim}

\smallskip
{\bf\noindent Projects vs. repositories}.
We made the assumption that each repository represents a project, but this is not necessarily true. A concrete example of this is \github{fraunhofer-iem/mechatronicuml} which was migrated
from a mono-repository to a multi-repository. This poses a limitation on our method since it does not handle well references across repositories in the same project. In the graph, this is reflected
as broken links (i.e., an edge pointing to an expected artefact which is not present in the repository). Nevertheless, such particular type of broken links could be fixed in a post-processing phase
applying heuristics (e.g., if there is a file path \code{../../sibling-repository/metamodel.ecore} it is likely a cross-repository reference).
%is that projects are split into several repositories.

\smallskip
{\bf\noindent Duplicate finding}. The deduplication algorithm depends on parameters which are set heuristically (see~\cite{allamanis2019adverse}). Also, we have applied adaptations of the algorithm
for XMI serialized artefacts (e.g., Ecore, Henshin, etc.) following ~\cite{lopez2022machine}. However, more studies are needed about the best way to find duplicates in modelling artefacts.
In particular, we have noticed that the algorithm sometimes do not fully capture synthetic duplicates, like the ones done for mutation experiments. 
% Citar DeLaVega - TOSEM

\smallskip
{\bf\noindent Temporal dimension}. The mega-model is computed on a specific snapshot of the cloned repositories (in particular, 2026/03/04). It is technically possible to have historical
views of the mega-model, but it requires checking out older branchs of each repository (which is fast) and then running the mega-model computation process again (which is slower).
To address this we would like to optimize the pipeline to support the efficient computation of past mega-models in order to study the technology evolution.
In any case, the mega-model do support some time-dependent analysis since each artefact is annotated with the creation and last update timestamps. This already allows
interesting applications like finding copy relationships between artefacts (that we use to compute the inter-project graph) or analysing the usage evolution of artefact types.

\subsection{Assessment}
In the following we realize an assessment of the dataset. For space reasons, the quantitative part
is limited to some aggregated metrics (Tables~\ref{tab:artifactTypes} and ~\ref{tab:mega_artefacts}).
Nevertheless, the online tool and the dataset package provide additional metrics.
%In the following we realize a quantiative and qualitative evaluation of the dataset, in particular focusing in the mega-model.

\smallskip
{\noindent\bf Artefacts}.
The final mega-model consists of \totalmderepos repositories and \totalartefacts MDE artefacts of 12 types. Table~\ref{tab:artifactTypes} shows the
distribution. Most of the artefacts are Ecore meta-models. There are also a large number of Henshin files but we have found that most of them are experimental
data and mutants (90\% of the Henshin artefacts belong to just 10 research projects). To understand better the how artefacts are distributed, the fourth column of Table~\ref{tab:artifactTypes} (Project ocurrences)
shows the percentage of projects with at least one artefact of the given type. In other words, it reflects the spread of each type of artefact in practice, regardless of its number of files.
As expected, Ecore is used in the majority of projects since, in MDE, all artefacts revolve around a meta-model.
Emftext and Henshin seems to be marginally used. Emftext do not receive updates since 7 years ago while Henshin seems to be mostly used in research projects.
A notable case is Xtext since it is used in about 40\% of the projects.

\smallskip
{\noindent\bf Duplication}.
To analyse duplication, we have counted the number of artefacts that are unique (see fifth column of Table~\ref{tab:artifactTypes}) in the sense that a) no duplicate is found by our algorithm
or b) if several artefacts are deemed as duplicated we count them as only one. Most of the Ecore artefacts are duplicates. Some of the largest duplication groups corresponde to well known meta-models
like \code{ATL.ecore}, \code{p2.ecore} (Eclipse updates plataform), \code{XML.ecore}, etc. which are copied from project to project for different purposes like testing, building datasets and especially
creating new transformations or providing a concrete syntax (see the running example, \code{uppaal.odesign} provides a concrete syntax to the copied \code{uppaal.ecore}).
Except for Henshin, the amount of duplication in the rest of the artefacts is less pronounced. In ATL, a typical source of duplication is
copy-paste-modify of transformations from the ATL Zoo. The amount of duplication Epsilon artefacts is smaller than similar artefacts like ATL. We hypothesize that it might be because they are used
for building actual modelling environments and less for doing research experiments or prototypes (which tend to reuse past data).

Therefore, the fact we have  have found a large amount of duplication all types of artefacts suggests that build management systems (akin to Maven) are in great need in the MDE community~\cite{sanchez2020build}.

%and there are little duplication of its artefacts.

\smallskip
{\noindent\bf Mega-model nodes and relationships}.
Table~\ref{tab:mega_artefacts} summarizes the recovered data at the mega-model level. There are more than 70,000 nodes.
An important aspect to consider is that the mega-model aggregates duplicate elements in a single node and therefore the number of nodes
is smaller than the number of processed artefacts. 
About 37\% of the nodes are isolated and no dependencies were derived from them.
The number of edges is about 77,000, with an average degree of 1,73 edges per node (excluding isolated nodes).
There are several causes for node isolation. In particular, most of the isolated nodes are Ecore meta-models (20,565 nodes). When an Ecore meta-model appears as an isolated node the causes can be:
(1) the references to the meta-model are found in a file whose technology we do not support, 2) there is a bug in an inspector,
3) the meta-model is referenced in Java code (which we do not consider) or 4) it is really an isolated model (typically this happens in repositories which belong to experiments or datasets).
We have found cases of (1) like \github{NCIP/lexevs} which applies XSLT transformations to XMI files directly.
There are also several cases of 3) of increasing complexity: from simple transformation launchers that configure a meta-model via its URI or file path and could be analysed heuristically building a dedicated
static analyser, to projects that load meta-models reflectively and therefore cannot be handled. % Cuales? NUIM-FM/asmig NicolasRouquette/qvto/tests
It is not possible to determine exactly the cause for a node to be isolated (whether it is a false negative or not) without a complete manual inspection, but we manually checked a large number of cases and
most of them are datasets or meta-models referenced from Java files.
We have also computed the number of connected components in the graph (see \textsc{\#components}). Each component indicates a cluster of related artefacts.
There is a very large component (29,239 nodes), a few more with 100 to 200 elements, and then many small components.

% Los que sí que identifico son reales porque se comprueba que el artefacto exista, en principio no hay falsos positivos, pero sí puede haber falsos negativos

%There are no false positives for edges of type ``exist'', then for edges of type heuristic there might be false positives but it is not possible to ensure check the actual validity since it depends on the original intentention of the developer.

%Many of the big groups correspond to test cases and experiments. For instance,
%Factories.ecore is group 
%antoniogarmendia/capone-graphical-pl/capone-examples/factories.pl.visualization/Factories.ecore

% artefacts by extension 
%select replace(name, rtrim(name, replace(name, '.', '' ) ), '') as ext, count(*) from artefacts where type is not 'ecore' group by ext;

%{\bf TODO: hablar del método para recuperar la información, por ejemplo, si ha sido con ant...}

\smallskip
{\noindent\bf Project relationships}. Table~\ref{tab:mega_artefacts} shows some statistics about the relationships identified using the inter-project graph.
The number of isolated projects is high (60\%) which means that many projects are build from scratch. Given the amount of duplication at the artefact level,
this means that project that reuse artefacts, tend to copy many of them. The average node degree is 1.74 which is in line with artefact connectivity.
There are very few connected components at the inter-project level, partly because there is a very large component with 2,681 projects.
In the future, we would like to have a more precise analysis trying to identify main hubs of the network.

%The connectivity of the inter-project graph is very high (number of edges . There reason is that the assumption that a project is related to other if it contains a duplicated element from another project reveals is too broad.
%This is because in a duplication group (i.e., a group of artefacts deemed as duplicates) we do not know which one is the original
%One way to address this issue would be to analyse the creation date of each artefact in the duplication group, but level temporal analysis is out of the scope of this paper.
%The inspection of graph reveals that there is a very large connected component (with 4301 projects) due to the fact just explained. Then, the rest of the components contain less 4 or less projects.
%
%The connectivity of the inter-project graph is very high (number of edges . There reason is that the assumption that a project is related to other if it contains a duplicated element from another project reveals is too broad.
%The inspection of graph reveals that there is a very large connected component (with 4301 projects) due to the fact just explained. Then, the rest of the components contain less 4 or less projects.

% Todo el tema de los repositorios...

% Graph
% Project
% Duplication

\begin{table}[]
  \caption{Graph statistics at the mega-model and project levels.}
  \label{tab:mega_artefacts}
  \begin{tabularx}{\columnwidth}{lr||Xr}
    \multicolumn{2}{c}{\textsc{Mega-model metrics}}     & \multicolumn{2}{c}{\textsc{Project metrics}} \\ 
    \#nodes & 70,645                  & \#projects & 7,436 \\
    \#isolated  & 26,304 (37\%)  & \#isolated  & 4,478 (60\%) \\    
    \#edges & 77,933                  & \#edges & 5,179 \\ 
    avg. degree & 1.76              & avg. degree & 1.75 \\
    \#components & 4,159       & \#components & 113 \\
    
%50|BUILTIN
%324263|EXISTS
%1250|GENERATED
%1752|HEURISTIC
%293|MISSING
%1183|UNEXPECTED
%1569|UNRESOLVED

    \toprule
  \end{tabularx}
\end{table}

% Poner estadísticas de esto

% Project statistics

% Poner estadísticas de número de paths heurísticos, por ejemplo

% We needed a list of built-in URIs. We use the list obtained from the Eclipse modeling edition which is...

% Decir o no que:
% We purposely ignore genmodel files...

%Limitations for the TrueIsolated strategy:

% See repo:
% 7ccsmmdd/turtle_dsl/TurtleTest/src/copy_turtles.etl
%We can't reason about generated meta-models by Xtext that are referenced by artefacts like ETL transformations. To solve this we should carefully order the analysis steps.

% See AGSNeditor/development/AGSN_sourcecode/AGSN/model/ECore2GMF.eol
% This uses implicitly Ecore and GMF which might not be properly matched by the heuristic based on footprints. What it is true is that we don't have evidence about how to run the transformation.

{\bf\noindent Building process}. The full process spans several days, primarily bounded by the time-intensive task of crawling and recovering artifact metadata from GitHub. Cloning the required repositories adds several hours and generates approximately 1.3 TB of data. Furthermore, identifying specific MDE artifacts involves traversing Git histories to locate both the oldest and newest commits (about 5 hours), while the final generation of the global mega-model and deduplication algorithms takes about 3 hours. To improve efficiency we want to explore newer GitHub API techniques~\cite{andre2026poolingh} and optimise some parts of the implementation.

%{\bf\noindent Building process}. Running the full process takes several days. The most time consuming part is crawling GitHub to extract relevant artefacts. Then, cloning the actual
%repositories also takes several hours, storing about 1.3 TB worth of data. Recovering repository information from GitHub also requires several hours. 
%As future work we would like to explore techniques like ~\cite{andre2026poolingh} to make the process of dealing with the GitHub API more efficient.
%The step of traversing the repositories to identify the actual artefacts takes around 5 hours because it needs to look up the oldest and newer commits for each MDE artefact individually.
%Finally, computing the global mega-model including running the duplication algorithms takes around 3 hours. % We currently keep a copy of the repositories...

%{\bf\noindent Usefulness.} Aquí se puede hablar de la experiencia personal

%The process of building the dataset has spanned four years (since 2022). 

\subsection{Potential usages of the dataset}\label{sec:usages}
In this section we discuss the usefulness of our work in terms of how the community could profit from it.

% Mencionar entrenar modelos de IA grandes 
% Mencionar replicar estudios previos o actualizarlos

At the {\bf repository level} the raw dataset could be used to answer questions about the maturity and health of the MDE projects. %, also in constrast to non-modelling projects.
This includes questions like: \textit{What is the engineering maturity of MDE projects?}, \textit{Are domain tools more or less mature than meta-tools?} or
\textit{What is the survival rate of MDE projects?}. To answer these questions techniques to compute engineered-project scoring (CI, contributors, commit history, issues, license) needs to be adapted to MDE~\cite{munaiah2017curating},
then classify projects into meta-tools and domain-tools (e.g., using ML techniques and datasets like ModelSet~\cite{lopez2022modelset}) and study metrics like commit distribution.

Another line of work is to study technology combinations in depth, answering questions like {\em Which MDE technologies are most commonly co-used within a single project?},
{\em Do projects combine several technologies?} or {\em How preferences about MDE technologies have changed over time?}.

At the {\bf mega-model level} the dataset could be used to analyse several aspects of how MDE is used in practice which are not well known yet. For instance, about network structures it would be interesting to answer: {\em Which are the central hubs?} and {\em What type of connected components can be identified?}. Also, it is possible to analyse structural patterns (motifs) like \code{Xtext $\to$ Metamodel $\to$ ATL} in order to discover
\textit{what are the most frequent structural motifs in the global mega-model}. Another important topic in MDE is reuse.
%Many techniques have been proposed to reuse MDE artefacts~\cite{kusel2015reuse} but they do not seem to have been applied in practice.
It would be interesting to answer questions like \textit{For which application domains are copy-paste more prevalent?} and devise methods to identify duplication causes: \textit{Do duplicates follow a ``provenance'' tree (one original,
  many copies) or a web (multiple independent derivations)?} and \textit{Is there evidence of unintentional copy-paste vs. intentional forking?}

The dataset can also be used to improve empirical evaluations of MDE tools. For instance, in~\cite{cuadrado2018anatlyzer} a static analyser is used to detect errors in 100 model ATL transformations curated in a repository. The study required to manually annotate each transformation with its meta-model. Now, this type of study could now be done at a larger scale (or applied to other technologies like Epsilon) because our system automatically perform this recovery and seamlessly provides a dataset of thousands of transformations.

Another line of work is to train or fine-tune AI models specific of MDE. The fact that the artefacts are already deduplicated would facilitate this task. In addition, the recorded relationships
can also be useful to enrich the training data.

% In this manuscript, we have only scratched the surface of the mega-model...
In this manuscript, for space reasons, we cannot further evaluate the contents of the mega-model or explore some of the research questions.
Nevertheless, we believe our contribution can be an important resource for the modelling community either to address the mentioned applications or others that we have not foreseen.

% Poner una tabla con las preguntas y con las featuers del dataset que ayudan a esto?

%Our proposal is also useful to enable the (semi-)automatic fixing of MDE projects. This could open a new research line similar to fixing builds in SE~\cite{fixing_builds}.

%Many times, using an MDE approach requires building a modellign tool for a particular project. This implies creating a meta-model, a concrete syntax (e.g., graphical, textual), plus generators, validators, etc.
%However, the time spent building it needs to be amortized across may projects. In the open source domain this can be more complicated since it is like building a new ``programming language for a very restricted domain'' which requires to have success, a community, etc.

%\section{Applying the dataset: Research Questions}\label{sec:questions}
%\input{research_questions}

%This is proably out of the scope of the paper
%\section{Analysis of MDE projects}
%\input{engineered_projects}

\section{Related work}\label{sec:related}

Several works have mined software repositories to perform empirical studies and understand the dynamics of software projects and artefacts.
For instance, Decan et al.~\cite{DBLP:journals/ese/DecanMG19} studied 830K packages from several package managers (e.g., Cargo, CPAN, CRAN or npm) to understand the evolution of dependencies among them.
%Also Decan et al.~\cite{DBLP:journals/tse/DecanM21} collected more than 900K packages to study the use of semantic versioning in package mangament systems.
%In the web engineering field, Di Lauro et al.~\cite{DBLP:conf/icwe/LauroSP21} have studied more 4.6K OpenAPI description to understand common evolution behaviors. 
In the Java ecosystem is common to analyse how projects use Maven dependencies. For instance, in ~\cite{sakib2025understanding} Maven packages have been analysed taking into account their popularity using information from GitHub,
while in \cite{suwanachote2025evolution} the focus is on the growth of unused dependencies as projects evolves. These types of empirical analysis have not been possible so far in the MDE
ecosystem. Our mega-model can play the role of the dependency information available in online package managers and enable these types of analysis.

% TODO: Encontrar algún trabajo más relacionado con la forma en que se hace aquí
In the modelling field some works have analysed open source projects to study the usage of specific types of MDE artefacts.
At the model level, UML and BPMN has been widely studied. In~\cite{heinze2020mining} around 8,000 BPMN models (after removing exact duplicates) were mined from GitHub and analysed to find violations of syntax and semantics.
Similarly, 20,000 BPMN models are mined and analysed in~\cite{compagnucci2021trends}. The latest effort to analyse BPMN~\cite{saeedi2025empirical} have analysed about 5,000 repositories and 25,000 models, finding that 90\% are clones. 
Regarding UML, a large dataset of UML models was made available in~\cite{hebig2016quest}, although the dataset is not actually analysed to check the validity of the artefacts.
%A retrospective analysis of the impact of the dataset~\cite{robles2023reflection}
More recently, in ~\cite{romeo2025uml} analyse the usage of UML models in GitHub. In our work, we have not included model artefacts, but we can observe large hubs around the UML and BPMN meta-models which highlights the
interest of these type of models.
%Therefore, as future work we plan to include in the mega-model instance level relationships for models.
% También: UML Usage in Open Source Software Development: A Field Study.

%Several works have crawled and analysed different types of MDE artefacts from public sources.
Other works have focused on the meta-level.
In \cite{zhang2026development} an empirical investigation about Xtext usage in GitHub is performed, which comprised manual classification of the repositories, locating DSL programs within the repositories
and identifying co-evolution issues across versions. 
In~\cite{babur2024language}, 32,832 unique Ecore meta-models were crawled  from GitHub, GHTorrent and the Software Heritage Dataset. The authors performed a deduplication phase but only by computing hashes. Then,
an analysis of how the Ecore meta-metamodel is used is performed (i.e., how the Ecore elements are used to create meta-models). It is also worth noting that~\cite{babur2024language} shows that the majority of public EMF meta-models reside in GitHub. A recent work have studied the causes of duplication for Ecore meta-models in public projects~\cite{de2025analysis}. Our dataset would enable replicating this study and extending it to a wider range of artefact types.
In~\cite{mengerink2019empowering} a large corpus of OCL expressions is contributed and analysed.
The MAR search engine crawled about 500,000 different types of models, including Ecore and Xtext files~\cite{lopez2022efficient}, however it does not identify relationships between the artefacts. We have used its list of crawled repositories to enhance our dataset.

%See Section 5.2.3.
% Comparar cuantos repos y artefactos encuentran

%The GitHub search method (see page 14) is based on downloading individual files. We download complete repositories.
%We have not included the repositories listed here since they are from before 2020, and thus already considered by MAR.
%(https://obabur.win.tue.nl/publications/EMSE22/)
% The OCL dataset!!

In terms of the method to recover artefact depedencies, our artefact graph recovery mechanism is partially inspired by \cite{di2020understanding}, but we add new support for broken dependencies through heuristics and cover many more file types. Moreover, the scope of this work is broader since we build a global mega-model and contributes an actual dataset. Also, our artefact graphs are an scalable way to recover information and even make the process parallelizable.
In the same line, AMINO~\cite{di2024amino} is a tool to discover relationships between modelling artefacts and to compute quality metrics. AMINO is not intended to be applied at scale and it only supports a small number of artefact types, but the proposed quality metrics could be applied to complement our dataset.
Tairas and Cabot~\cite{tairas2015corpus} presents a method to analyse DSL usage, including clone detection and analysis. These techniques could be applied to our
dataset to derive new knowledge about MDE languages usage. % including trends of the usage of their constructs

There are few works that have analysed complete projects. The most notable exception is the work by C\'anovas Izquierdo et al.~\cite{izquierdo2017empirical} which analysed the maturity level of modelling projects hosted by Eclipse.
They show that modeling projects are slightly less mature than non-modelling projects.
A recent work have analysed the historical trend of model transformation language usage~\cite{de2026have}. Our work can be used to complement such analysis and shed more light
on the causes of the decline of model transformation technology.

% Papers the ralph laemmel

%\cite{heinz2020reproducible} proposes the idea of ``technology models'' to summarize how technologies are used. Such models are represented 

%% See 101 companies effort???

%Modeling the linguistic architecture of software products%

%Tales of ... Xtext

%The idea of using mega-models to describe the linguistic architecture of software systems has been proposed in\cite{favre2012modeling}.
%Specific relationships are defined using the MegaL language. In our case, the relationships are simpler and harcoded in the mega-model.

% Paper TOTEM con L'Aquila

%See Mining BPMN processes on GitHub for tool validation and development. heinze2020mining  => it seems that they do deduplication

%Regarding strategies to mine GitHub, a recent work have proposed a tool to improve API usage\cite{andre2026poolingh}

\section{Conclusion}\label{sec:conclusion}

In this paper we have presented a method to construct a global mega-model from public MDE artefacts. We have considered
12 different artefact types mined from GitHub, encompassing meta-models, transformations, concrete syntax and code generators.
The mega-model represents relationships between artefacts at different levels: individual files, project and global levels.
To the best of our knowlege, this is the largest dataset of MDE artefacts and the only one describing artefact relationships.
% Habría que concluir hablando de modelling environments

As future work we plan to continue enhancing the dataset by improving the artefact inspectors (e.g., recover dependencies from Java files)
and also including instance level artefacts by crawling models like UML and BPMN models.
We also would like to address the temporal dimension by providing a sequence of mega-models captured at fixed timestamps.
Regarding the use of the dataset, we plan to start addressing some of the research questions, in particular those related to network analysis to identify
MDE usage patterns. We also want to experiment training AI models with the dataset.

%\begin{acks}
%To Robert, for the bagels and explaining CMYK and color spaces.
%\end{acks}

\bibliographystyle{ACM-Reference-Format}
\bibliography{article}

@article{liebel2018model,
  title={Model-based engineering in the embedded systems domain: an industrial survey on the state-of-practice},
  author={Liebel, Grischa and Marko, Nadja and Tichy, Matthias and Leitner, Andrea and Hansson, J{\"o}rgen},
  journal={Software \& Systems Modeling},
  volume={17},
  number={1},
  pages={91--113},
  year={2018},
  publisher={Springer}
}

@article{munaiah2017curating,
  title={Curating github for engineered software projects},
  author={Munaiah, Nuthan and Kroh, Steven and Cabrey, Craig and Nagappan, Meiyappan},
  journal={Empirical Software Engineering},
  volume={22},
  number={6},
  pages={3219--3253},
  year={2017},
  publisher={Springer}
}

@inproceedings{sanchez2020build,
  title={To build, or not to build: ModelFlow, a build solution for MDE projects},
  author={S{\'a}nchez, Beatriz and Kolovos, Dimitris and Paige, Richard},
  booktitle={Proceedings of the 23rd ACM/IEEE International Conference on Model Driven Engineering Languages and Systems},
  pages={1--11},
  year={2020}
}

@article{mengerink2019empowering,
  title={Empowering OCL research: a large-scale corpus of open-source data from GitHub},
  author={Mengerink, Josh GM and Noten, Jeroen and Serebrenik, Alexander},
  journal={Empirical Software Engineering},
  volume={24},
  number={3},
  pages={1574--1609},
  year={2019},
  publisher={Springer}
}

@inproceedings{suwanachote2025evolution,
  title={On the evolution of unused dependencies in Java project releases: an empirical study},
  author={Suwanachote, Nabhan and Shakizada, Yagut and Kashiwa, Yutaro and Lin, Bin and Iida, Hajimu},
  booktitle={2025 IEEE/ACM 22nd International Conference on Mining Software Repositories (MSR)},
  pages={324--328},
  year={2025},
  organization={IEEE}
}

@inproceedings{sakib2025understanding,
  title={Understanding the popularity of packages in Maven ecosystem},
  author={Sakib, Sadman Jashim and Asaduzzaman, Muhammad and Bright, Curtis and Morgan, Cole},
  booktitle={2025 IEEE/ACM 22nd International Conference on Mining Software Repositories (MSR)},
  pages={364--368},
  year={2025},
  organization={IEEE}
}

@article{zhang2026development,
  title={Development and evolution of Xtext-based DSLs on GitHub: an empirical investigation},
  author={Zhang, Weixing and Str{\"u}ber, Daniel and Hebig, Regina},
  journal={Empirical Software Engineering},
  volume={31},
  number={3},
  pages={48},
  year={2026},
  publisher={Springer}
}

@article{lopez2022modelset,
  title={ModelSet: a dataset for machine learning in model-driven engineering},
  author={L{\'o}pez, Jos{\'e} Antonio Hern{\'a}ndez and Canovas Izquierdo, Javier Luis and Cuadrado, Jes{\'u}s S{\'a}nchez},
  journal={Software and Systems Modeling},
  volume={21},
  number={3},
  pages={967--986},
  year={2022},
  publisher={Springer}
}

@inproceedings{hebig2016quest,
  title={The quest for open source projects that use UML: mining GitHub},
  author={Hebig, Regina and Quang, Truong Ho and Chaudron, Michel RV and Robles, Gregorio and Fernandez, Miguel Angel},
  booktitle={Proceedings of the ACM/IEEE 19th international conference on model driven engineering languages and systems},
  pages={173--183},
  year={2016}
}

@inproceedings{compagnucci2021trends,
  title={Trends on the usage of BPMN 2.0 from publicly available repositories},
  author={Compagnucci, Ivan and Corradini, Flavio and Fornari, Fabrizio and Re, Barbara},
  booktitle={International Conference on Business Informatics Research},
  pages={84--99},
  year={2021},
  organization={Springer}
}

@article{di2024amino,
  title={AMINO: A quality assessment framework for modeling ecosystems},
  author={Di Ruscio, Davide and Iovino, Ludovico and Pierantonio, Alfonso},
  journal={Journal of Software: Evolution and Process},
  volume={36},
  number={5},
  pages={e2603},
  year={2024},
  publisher={Wiley Online Library}
}

@inproceedings{heinze2020mining,
  title={Mining BPMN Processes on GitHub for tool validation and development},
  author={Heinze, Thomas S and Stefanko, Viktor and Amme, Wolfram},
  booktitle={International Conference on Business Process Modeling, Development and Support},
  pages={193--208},
  year={2020},
  organization={Springer}
}

@article{tairas2015corpus,
  title={Corpus-based analysis of domain-specific languages},
  author={Tairas, Robert and Cabot, Jordi},
  journal={Software \& Systems Modeling},
  volume={14},
  number={2},
  pages={889--904},
  year={2015},
  publisher={Springer}
}

@inproceedings{andre2026poolingh,
  title={PoolinGH: Fast, Efficient, and Robust GitHub Repository Mining},
  author={Andr{\'e}, Maxime and Raglianti, Marco and Serbout, Souhaila and Cleve, Anthony and Lanza, Michele},
  booktitle={Proceedings of the 23rd International Mining Software Repositories Conference (MSR 2026): Data and Tool Showcase Track},
  year={2026},
  organization={ACM Press}
}

@article{de2026have,
  title={Have model transformation languages failed? On the rise, fall and revival of model transformation languages},
  author={de Lara, Juan and Guerra, Esther and Cuadrado, Jes{\'u}s S{\'a}nchez},
  journal={Software and Systems Modeling},
  pages={1--15},
  year={2026},
  publisher={Springer}
}

@article{zolotas2020bridging,
  title={Bridging proprietary modelling and open-source model management tools: the case of PTC Integrity Modeller and Epsilon},
  author={Zolotas, Athanasios and Hoyos Rodriguez, Horacio and Hutchesson, Stuart and Sanchez Pina, Beatriz and Grigg, Alan and Li, Mole and Kolovos, Dimitrios S and Paige, Richard F},
  journal={Software and Systems Modeling},
  volume={19},
  number={1},
  pages={17--38},
  year={2020},
  publisher={Springer}
}

@inproceedings{lopez2022machine,
  title={Machine learning methods for model classification: a comparative study},
  author={L{\'o}pez, Jos{\'e} Antonio Hern{\'a}ndez and Rubei, Riccardo and Cuadrado, Jes{\'u}s S{\'a}nchez and Di Ruscio, Davide},
  booktitle={Proceedings of the 25th International Conference on Model Driven Engineering Languages and Systems},
  pages={165--175},
  year={2022}
}

@inproceedings{allamanis2019adverse,
  title={The adverse effects of code duplication in machine learning models of code},
  author={Allamanis, Miltiadis},
  booktitle={Proceedings of the 2019 ACM SIGPLAN international symposium on new ideas, new paradigms, and reflections on programming and software},
  pages={143--153},
  year={2019}
}

@inproceedings{bezivin2004need,
  title={On the need for megamodels},
  author={B{\'e}zivin, Jean and Jouault, Fr{\'e}d{\'e}ric and Valduriez, Patrick},
  booktitle={proceedings of the OOPSLA/GPCE: best practices for model-driven software development workshop, 19th Annual ACM conference on object-oriented programming, systems, languages, and applications},
  pages={1--9},
  year={2004}
}

@article{lopez2022efficient,
  title={An efficient and scalable search engine for models},
  author={L{\'o}pez, Jos{\'e} Antonio Hern{\'a}ndez and Cuadrado, Jes{\'u}s S{\'a}nchez},
  journal={Software and Systems Modeling},
  volume={21},
  number={5},
  pages={1715--1737},
  year={2022},
  publisher={Springer}
}

@inproceedings{cuadrado2018anatlyzer,
  title={Anatlyzer: An advanced ide for atl model transformations},
  author={Cuadrado, Jes{\'u}s S{\'a}nchez and Guerra, Esther and de Lara, Juan},
  booktitle={Proceedings of the 40th international conference on software engineering: Companion proceeedings},
  pages={85--88},
  year={2018}
}

@article{lara2019automated,
  title={Automated reuse of model transformations through typing requirements models},
  author={Lara, Juan De and Guerra, Esther and Ruscio, Davide Di and Rocco, Juri Di and Cuadrado, Jes{\'u}s S{\'a}nchez and Iovino, Ludovico and Pierantonio, Alfonso},
  journal={ACM Transactions on Software Engineering and Methodology (TOSEM)},
  volume={28},
  number={4},
  pages={1--62},
  year={2019},
  publisher={ACM New York, NY, USA}
}

@inproceedings{romeo2025uml,
  title={UML is back. Or is it? Investigating the past, present, and future of UML in open source software},
  author={Romeo, Joseph and Raglianti, Marco and Nagy, Csaba and Lanza, Michele},
  booktitle={2025 IEEE/ACM 47th International Conference on Software Engineering (ICSE)},
  pages={2342--2354},
  year={2025},
  organization={IEEE}
}

@article{de2025analysis,
  title={Analysis of EMF meta-model duplication in open-source repositories},
  author={de la Vega, Alfonso and Hern{\'a}ndez L{\'o}pez, Jos{\'e} Antonio},
  journal={ACM Transactions on Software Engineering and Methodology},
  year={2025},
  publisher={ACM New York, NY}
}

@article{saeedi2025empirical,
  title={An empirical study of business process models and model clones on GitHub},
  author={Saeedi Nikoo, Mahdi and Kochanthara, Sangeeth and Babur, {\"O}nder and van den Brand, Mark},
  journal={Empirical Software Engineering},
  volume={30},
  number={2},
  pages={48},
  year={2025},
  publisher={Springer}
}

@inproceedings{hutchinson2011empirical,
  title={Empirical assessment of MDE in industry},
  author={Hutchinson, John and Whittle, Jon and Rouncefield, Mark and Kristoffersen, Steinar},
  booktitle={Proceedings of the 33rd international conference on software engineering},
  pages={471--480},
  year={2011}
}

@inproceedings{izquierdo2017empirical,
  title={An empirical study on the maturity of the eclipse modeling ecosystem},
  author={Izquierdo, Javier Luis C{\'a}novas and Cosentino, Valerio and Cabot, Jordi},
  booktitle={2017 ACM/IEEE 20th International Conference on Model Driven Engineering Languages and Systems (MODELS)},
  pages={292--302},
  year={2017},
  organization={IEEE}
}

@article{DBLP:journals/ese/DecanMG19,
  author       = {Alexandre Decan and Tom Mens and Philippe Grosjean},
  title        = {An empirical comparison of dependency network evolution in seven software packaging ecosystems},
  journal      = {Empir. Softw. Eng.},
  volume       = {24},
  number       = {1},
  pages        = {381--416},
  year         = {2019}
}

@article{DBLP:journals/tse/DecanM21,
  author       = {Alexandre Decan and Tom Mens},
  title        = {What Do Package Dependencies Tell Us About Semantic Versioning?},
  journal      = {{IEEE} Trans. Software Eng.},
  volume       = {47},
  number       = {6},
  pages        = {1226--1240},
  year         = {2021}
}

@article{di2020understanding,
  title={Understanding MDE projects: megamodels to the rescue for architecture recovery},
  author={Di Rocco, Juri and Di Ruscio, Davide and H{\"a}rtel, Johannes and Iovino, Ludovico and L{\"a}mmel, Ralf and Pierantonio, Alfonso},
  journal={Software and Systems Modeling},
  volume={19},
  pages={401--423},
  year={2020},
  publisher={Springer}
}

@article{babur2024language,
  title={Language usage analysis for EMF metamodels on GitHub},
  author={Babur, {\"O}nder and Constantinou, Eleni and Serebrenik, Alexander},
  journal={Empirical Software Engineering},
  volume={29},
  number={1},
  pages={23},
  year={2024},
  publisher={Springer}
}

%\appendix

\end{document}